# State and Parameter Estimation Based on Filtered Transformation for a Class of Second-Order Systems

Mehdi Tavan, Kamel Sabahi, and Saeid Hoseinzadeh

*Abstract*— This paper addresses the problem of state and parameter estimation for a class of second-order systems with single output. A new filtered transformation is proposed for the system via dynamic vector and matrix. In this method, the dynamics of the vector and matrix are derived by immersion and invariance technique such that the state estimation condition is guaranteed. Compared to the classical approaches that persistency of excitation (PE) condition is required for parameter convergence, the proposed method needs a weaker one, so called non-square-integrability condition, in the transformation via dynamic matrix. Simulation results are concluded for a class of regressors which are not PE but satisfy the new condition.

## I. INTRODUCTION

The *persistency of excitation* (PE) condition on the regressor is one of the main tools to ensure convergence in parameter estimators for identification purposes. Recently, a new procedure has introduced in [1] to design the parameter estimators in which the PE condition is replaced with a weaker one, called non-square-integrability condition. In this procedure some dynamic regressors are designed to guarantee the new condition for the determinant of an extended regressor. Applications of this method for some electrical examples show better transient and robust performance properties [2, 3]. The use of dynamic regressors can be found for adaptive observer design via filtered transformation in [4] and the references cited therein. Besides the PE condition for parameter convergence, an additional disadvantage is *overparameterisation* in such a design. The *immersion and invariance* (I&I) based filtered transformation introduced in [5] reduces the order of the observer with elimination of output injection error term in update law. Application of this method for DC-DC boost converter is evaluated via experiments in [6]. The filtered transformations in [4, 5] are formed via a dynamic vector which yields to a descent algorithm with *singular* descent rate that is at the core of convergence problem in parameter estimation [7].

In this paper a filtered transformation is introduced via dynamic matrix for a class of second-order systems with single-output. The main contribution of this paper is to propose an estimator based on the new transformation, which asymptotically estimates unavailable state and unknown parameters. The parameter convergence is ensured under the non-square-integrability condition. Fulfilling this condition is discussed for the estimator. Moreover, it is shown that the estimator provides an arbitrary convergence speed. Also simulation results are concluded for a class of regressors which are not PE but satisfy the new condition.

## II. PROBLEM STATEMENT AND PRELIMINARY

Consider a class of single-output nonlinear systems that can be described in suitable co-ordinates by equations of the form

$$\dot{y} = f(y,t)x + g_0(y,t), \qquad (1)$$

$$\dot{x} = g_1(y,t) + \varphi^\top(y,t)\theta, \qquad (2)$$

where $y \in \mathbb{R}$ is the measurable state, $x \in \mathbb{R}$ is the unmeasurable state, $\theta \in \mathbb{R}^q$ is a constant vector of unknown parameters, $f(.), g_0(.), g_1(.)$, and $\varphi(.)$ are known mappings. As mentioned in Introduction, our objective consists of estimating $x$ and $\theta$. As shown in [6], the average model of some popular switch converters, such as DC-DC boost, buck and buck-boost, can be represented in the form of above equations.

For the system, the unmeasurable state $x$ can be assumed to be generated by a parameter dependent change of coordinate $p \in \mathbb{R}$ [4]

$$\begin{aligned} x &= p + \mu^\top \theta \\ &= [1 \quad \mu^\top]\eta, \end{aligned} \qquad (3)$$

where $\mu(t): \mathbb{R}_+ \to \mathbb{R}^q$ is an auxiliary dynamic vector, which its dynamics are to be defined, and $\eta = \text{col}[p, \theta]$ is the unavailable vector. From (2) and (3) the dynamics of $\eta$ can be written as

$$\dot{\eta} = A\eta + \begin{bmatrix} g_1 \\ o_q \end{bmatrix}, \qquad (4)$$

with

$$A = \begin{bmatrix} 0 & (\varphi - \dot{\mu})^\top \\ o_q & O_q \end{bmatrix},$$

where $O_q \in \mathbb{R}^{q \times q}$ and $o_q \in \mathbb{R}^q$ are the zero matrix and zero vector, respectively. Although using the vector filtered transformation $\mu(t)$ for such systems like in [5, 6] facilitate the state estimator design, the limitation imposed by the PE condition is still remained in the parameter estimation. To overcome this problem the matrix filtered transformation is introduced as

$$\begin{aligned} \chi &= \pi + M^\top \theta \\ &= [I_q \quad M^\top]\vartheta, \end{aligned} \qquad (5)$$

where $\chi = \iota x \in \mathbb{R}^q$, in which $\iota = \text{col}(1, \dots, 1) \in \mathbb{R}^q$, $\pi \in \mathbb{R}^q$, $I_q = \text{dig}(\iota) \in \mathbb{R}^{q \times q}$ is identity matrix, $M(t): \mathbb{R}_+ \to \mathbb{R}^{q \times q}$ is

M. Tavan and Saeid Hoseinzadeh are with the Department of Electrical Engineering, Mahmudabad Branch, Islamic Azad University, Mahmudabad, Iran (e-mail: m.tavan@srbiau.ac.ir).

K. Sabahi is with the Department of Electrical Engineering, Mamaghan Branch, Islamic Azad University, Mamaghan, Iran.

an auxiliary dynamic matrix, which its dynamics are to be defined, and $\vartheta = \text{col}[\pi, \theta]$ is the new unavailable vector. In turn, the dynamics of $\vartheta$ can be obtained as

$$\dot{\vartheta} = \Lambda \vartheta + \begin{bmatrix} \gamma_1 \\ 0_q \end{bmatrix}, \tag{6}$$

with

$$\Lambda = \begin{bmatrix} 0_q & (\Phi - \dot{M})^\top \\ 0_q & 0_q \end{bmatrix},$$

where $\gamma_1(y, t) = \iota g_1(y, t): \mathbb{R} \times \mathbb{R}_+ \to \mathbb{R}^q$ and $\Phi^\top(y, t) = \iota \varphi^\top(y, t): \mathbb{R} \times \mathbb{R}_+ \to \mathbb{R}^{q \times q}$.

As will become clear in the next section, the derivation of the $\mu(t)$-dynamics and the $M(t)$-dynamics follows the idea in [5, 6]. In the proposed approach, the advantages of I&I technique with input-output filtered transformation is combined to construct a proper immersion and an auxiliary dynamic filter for the estimation purpose.

### III. ESTIMATOR DESIGN

In this section, two classes of estimators are designed for the system (1), (2). Although, the first one facilitates estimator design using a filtered transformation via dynamic vector, it requires the regressor persistency of excitation for the parameter estimation. It is worth pointing out that the second one ensures parameter convergence with PE dismissal.

In order that the state $x$ be observable from the output $y$, the following assumption is imposed.

*Assumption 1:* The system (1), (2) is forward complete, i.e. trajectories exist for all $t \in \mathbb{R}_+$. The function $f(y, t)$ and its time derivative are bounded and $|f| \in \mathbb{R}_{>0}$ for all $y \in \mathbb{R}$ and $t \in \mathbb{R}_+$.

### A. Filtered Transformation via Dynamic Vector

This subsection is devoted to summarize the method proposed in [5, 6] with some modifications. The change of coordinate (3) and the dynamic equation (4) are considered for design methodology in this subsection. Following the I&I procedure [8] and [5, 6], define the *estimation error* as

$$z = \eta - \zeta - \beta(y, \mu), \tag{7}$$

where $\zeta \in \mathbb{R}^{1+q}$ is the estimator state, the dynamics of which is to be defined, and the mapping $\beta: \mathbb{R} \times \mathbb{R}_+ \to \mathbb{R}^{1+q}$ yet to be specified. To obtain the dynamics of $z$, differentiate (7), yielding

$$\dot{z} = \dot{\eta} - \dot{\zeta} - \frac{\partial \beta}{\partial y} \dot{y} - \frac{\partial \beta}{\partial \mu} \dot{\mu},$$

where $\dot{\eta}$ is given in (4), and the output dynamics $\dot{y}$ can be rewritten in term of $\eta$, as

$$\dot{y} = f(y, t)[1 \quad \mu^\top] \eta + g_0(y, t). \tag{8}$$

Let

$$\dot{\zeta} = -\left(\frac{\partial \beta}{\partial y} f [1 \quad \mu^\top] - A\right)(\zeta + \beta) + \begin{bmatrix} g_1 \\ 0_q \end{bmatrix} - \frac{\partial \beta}{\partial y} g_0 - \frac{\partial \beta}{\partial \mu} \dot{\mu}, \tag{9}$$

where $\dot{\mu}$ is to be defined. Replacing (9) in (7), yields

$$\dot{z} = -\left(\frac{\partial \beta}{\partial y} f [1 \quad \mu^\top] - A\right) z. \tag{10}$$

From (7) and the definition of $\eta = \text{col}[p, \theta]$, the estimation error can be decomposed into $z = \text{col}[z_1, z_2]$ with $z_1 = p - \zeta_1 - \beta_1$ and $z_2 = \theta - \zeta_2 - \beta_2$. Also, with respect to (3), the state estimation error can be expressed as $\bar{x} = z_1 + \mu^\top z_2$. The following proposition establishes that there exists the mapping $\beta$ and the dynamics of the filtered vector $\mu$ such that the state estimation error $\bar{x}$ converges to zero.

*Proposition 1:* Consider the system (10) under Assumption 1 with

$$\beta(y, \mu) = \begin{bmatrix} \beta_1(y) \\ \beta_2(y, \mu) \end{bmatrix} = \text{sgn}(f) \begin{bmatrix} 1 \\ \Gamma B \mu \end{bmatrix} k(y), \tag{11}$$

$$\dot{\mu} = -|f| \frac{\partial k}{\partial y} (I_q + B) \mu + \varphi(y, t), \tag{12}$$

where $B = B^\top \in \mathbb{R}^{q \times q}$ and $\Gamma = \Gamma^\top \in \mathbb{R}^{q \times q}$ are constant and positive definite matrices, $k(y): \mathbb{R} \to \mathbb{R}$ is a strictly increasing function with bounded derivative. Then the system (10) has a uniformly globally stable equilibrium at the origin, $z \in \mathcal{L}_\infty$ and $\bar{x} \in \mathcal{L}_2$. If, in addition, $\varphi(.)$ is bounded, then $\mu(t)$ is bounded and $\bar{x}$ converges to zero.

*Proof 1:* Notice that the derivation of the signum function $\text{sgn}(f)$ is zero, due to the observation restriction $f(.) \neq 0$. Substituting (11), (12) into (10) yields the error dynamics

$$\dot{z} = -\begin{bmatrix} \frac{\partial \beta_1}{\partial y} f & \left(\dot{\mu} + \frac{\partial \beta_1}{\partial y} f \mu - \varphi\right)^\top \\ \frac{\partial \beta_2}{\partial y} f & \frac{\partial \beta_2}{\partial y} f \mu^\top \end{bmatrix} z$$

$$= -|f| \frac{\partial k}{\partial y} \begin{bmatrix} 1 & -\mu^\top B \\ \Gamma B \mu & \Gamma B \mu \mu^\top \end{bmatrix} z, \tag{13}$$

where $|f| = f \text{sgn}(f)$ has been used to get the last equality. Consider now the following Lyapunov function candidate

$$V(z) = \frac{1}{2}(z_1^2 + z_2^\top \Gamma^{-1} z_2), \tag{14}$$

whose time-derivative along the trajectories of (13) satisfies

$$\dot{V} = -|f| \frac{\partial k}{\partial y} (z_1^2 + z_2^\top \mu \mu^\top B z_2)$$

$$\leq -|f| \frac{\partial k}{\partial y} (z_1^2 + c_0 (\mu^\top z_2)^2), \tag{15}$$

for some constant $c_0 \in \mathbb{R}_{>0}$, where the positive definite property of $B$ has been used to get the bound. The equations (14) and (15) imply that the system (13) has a uniformly globally stable equilibrium at the origin, $z \in \mathcal{L}_\infty$, $z_1 \in \mathcal{L}_2$, $\mu^\top z_2 \in \mathcal{L}_2$ and then $\bar{x} \in \mathcal{L}_2$. Note that the above holds for any $\varphi$. It is clear that, (12) is *input-to-state stable* dynamic, then $\mu(t)$ and its time derivative remains bounded for any bounded $\varphi$. Consequently, $\mu$, $f$, $\dot{f}$, and $\partial k/\partial y$ being bounded, $\dot{z}$ in (13) and then $\ddot{V}$ in (15) are bounded. Therefore, the convergence of $\bar{x} = z_1 + \mu^\top z_2$ to zero follows directly from Lemma 1 in [9]. ∎

*Remark 1:* If $\bar{x} = [1 \ \mu^\top] z$ converges to zero, an asymptotic estimate of $x$ is given by $\hat{x} = [1 \ \mu^\top](\zeta + \beta)$. As is well known ([10], Lemma B.2.3), the parameter estimation error $z_2$ converges to zero if the regressor vector $\mu(t)$ and its time derivative are uniformly bounded and the regressor satisfies PE condition. Notice that, the bounded condition is satisfied here for any bounded $\varphi$.

*Remark 2:* In comparison with the estimator proposed in [5, 6], the gain matrix $B$ is added to filter dynamics in (12) which can help to make the regressor *exciting* over a time sequence arbitrary.

*Remark 3:* In comparison with the mapping $\beta$ in [5, 6], the gain matrix $\Gamma$ is added in (11) and then in error dynamics (13). The system (13) is in the form of damped nonlinear oscillator in which the transient time can be improved arbitrary by reducing the overshot via $\Gamma$ tuning.

### B. Filtered Transformation via Dynamic Matrix

To dismissal the PE condition in parameter estimation, the design methodology in this subsection is based on the change of coordinate (5) and the dynamic equation (6). In the light of the proposed design procedure in the previous subsection, a general expression of $\beta(y,t)$ can be expressed by

$$z = \vartheta - \zeta - \beta(y,t), \tag{16}$$

$$\beta(y,t) = \text{sgn}(f) \begin{bmatrix} I_q \\ S(t) \end{bmatrix} \kappa(y), \tag{17}$$

where $S(t): \mathbb{R}_+ \to \mathbb{R}^{q \times q}$ is to be defined, and $\kappa(y) = \iota k(y): \mathbb{R} \to \mathbb{R}^q$ is a vector function in which, $k(y)$ is defined in Proposition 1. Notice that, the dynamics of the vector function $\kappa(y)$ can be obtained in term of $\vartheta$, as

$$\begin{aligned} \dot{\kappa} &= \frac{\partial k}{\partial y}\big(f(y,t)\chi + \gamma_0(y,t)\big) \\ &= \frac{\partial k}{\partial y}\big(f(y,t)[I_q \ M^\top]\vartheta + \gamma_0(y,t)\big), \end{aligned} \tag{18}$$

where $\gamma_0(y,t) = \iota g_0(y,t): \mathbb{R} \times \mathbb{R}_+ \to \mathbb{R}^q$. Now, by choosing the update laws as

$$\dot{\zeta} = -|f|\frac{\partial k}{\partial y}\begin{bmatrix} I_q & \psi^\top \\ S & SM^\top \end{bmatrix}(\zeta + \beta) + \begin{bmatrix} \gamma_1 \\ 0_q \end{bmatrix}$$
$$-\text{sgn}(f)\left(\frac{\partial k}{\partial y}\begin{bmatrix} I_q \\ S \end{bmatrix}\gamma_0 + \begin{bmatrix} 0_q \\ \dot{S} \end{bmatrix}\kappa(y)\right), \tag{19}$$

$$\dot{M}^\top = -|f|\frac{\partial k}{\partial y}(M^\top - \psi^\top) + \Phi^\top(y,t), \tag{20}$$

and with respect to (6), (16)-(18), the resulting error dynamics can be obtained as

$$\dot{z} = -|f|\frac{\partial k}{\partial y}\begin{bmatrix} I_q & \psi^\top \\ S & SM^\top \end{bmatrix} z. \tag{21}$$

From (16) and the definition of $\vartheta = \text{col}[\pi, \theta]$, the estimation error $z$ can be decomposed into $z_1 = \pi - \zeta_1 - \beta_1$ and $z_2 = \theta - \zeta_2 - \beta_2$. Also, with respect to (5), the state estimation error of $\chi$ can be stated as $\bar{\chi} = z_1 + M^\top z_2$. A choice of $S$ and $\psi$ which achieves the design objective is given in the following proposition.

*Proposition 2:* Consider the system (21) verifying Assumption 1, with

$$S(t) = -\Gamma\psi(t), \tag{22}$$

$$\psi(t) = -M(t)B, \tag{23}$$

where $\Gamma$, $B$ and $k(y)$ are chosen as in Proposition 1, in addition, suppose that the eigenvalues of $B$ are different. Then the system (21) has a uniformly globally stable equilibrium at the origin, $z \in \mathcal{L}_\infty$ and $\bar{\chi} \in \mathcal{L}_2$. Moreover, if $\varphi(.)$ is bounded, then $M(t)$ is bounded and $\bar{\chi}$ converges to zero. If in addition $\det(M) \notin \mathcal{L}_2$ (resp. $\det(M) \neq 0$), then $z_2$ converges to zero asymptotically (resp. exponentially).

*Proof 2:* Consider the following Lyapunov function candidate

$$V(z) = \tfrac{1}{2}(z_1^\top z_1 + z_2^\top \Gamma^{-1} z_2), \tag{24}$$

whose time-derivative along the trajectories of (21), is

$$\dot{V} = -\tfrac{1}{2}|f|\frac{\partial k}{\partial y} z^\top \begin{bmatrix} 2I_q & \psi^\top + S^\top \Gamma^{-1} \\ \Gamma^{-1}S + \psi & \Gamma^{-1}SM^\top + MS^\top\Gamma^{-1} \end{bmatrix} z. \tag{25}$$

Replacing (23) and (24) in (25), we obtain

$$\dot{V} = -|f|\frac{\partial k}{\partial y}(z_1^\top z_1 + z_2^\top MBM^\top z_2), \tag{26}$$

which implies that the system (21) has a uniformly globally stable equilibrium at the origin, $z \in \mathcal{L}_\infty$, $z_1 \in \mathcal{L}_2$, $M^\top z_2 \in \mathcal{L}_2$ and then $\bar{\chi} \in \mathcal{L}_2$. The above holds for any $\Phi = \varphi \iota^\top$. Inserting (23) and (24) into (20) yields

$$\dot{M}^\top = -|f|\frac{\partial k}{\partial y}(I_q + B)M^\top + \Phi^\top(y,t), \tag{27}$$

which is *input-to-state stable* with respect to perturbation $\Phi^\top(.)$. Therefore, the dynamic equation (26) with $\Phi = \varphi\iota^\top \in \mathcal{L}_\infty$ induces $M, \dot{M} \in \mathcal{L}_\infty$. On the other hand, for bounded $M$, $f$, $\dot{f}$, and $\partial k/\partial y$, boundedness of $\dot{z}$ and $\ddot{V}$ can be obtained from (21)-(23) and (26) respectively. As a result, the convergence of $M^\top z_2$ and then $\bar{\chi}$ to zero is guaranteed by Lemma 1 in [9]. Note now that

$$z_2^\top MBM^\top z_2 \geq c_1 z_2^\top MM^\top z_2, \tag{28}$$

for some $c_1 \in \mathbb{R}_{>0}$, where use has been made of positive definite property of $B$. For any $M(t) \in \mathbb{R}^{q \times q}$, the expression $M(t)M^\top(t)$ is symmetric, and its eigenvalues are real and nonnegative [11]. Also, if $\lambda_i(t)$ are the eigenvalues of $M(t)$, then $\lambda_i^2(t)$ are the eigenvalue of $M(t)M^\top(t)$, for all $i =$

$1, \ldots, q$. Let $\lambda_m^2(t)$ be the minimal value of $\lambda_i^2(t)$ for any $t \in \mathbb{R}_+$, then

$$M(t)M^\top(t) \geq \lambda_m^2(t) I_q. \tag{29}$$

On the other hand, it is well known the determinant of a square matrix is the product of all its eigenvalues. As a result, if $\det(M) \neq 0$ then $\lambda_m^2(t) \in \mathbb{R}_{>0}$. Now using (26)-(29) and doing some basic bounding, yields

$$\begin{aligned}\dot{V} &\leq -|f|\frac{\partial k}{\partial y}(z_1^\top z_1 + \lambda_m^2 z_2^\top z_2) \\ &\leq -c_2 \sigma(t) V,\end{aligned} \tag{30}$$

for some $c_2 \in \mathbb{R}_{>0}$ and $\sigma(t) = \min(1, \lambda_m^2)$. It is clear that exponential convergence of $V$, and then $z$ can be directly obtained from (30) due to the positiveness of $\sigma(t)$. Note now that for bounded $\lambda_i(t)$, if $\det(M) \notin \mathcal{L}_2$ then all eigenvalues $\lambda_i^2(t) \notin \mathcal{L}_2$ [1]. As a result, $\lambda_m^2(t), \sigma(t) \notin \mathcal{L}_2$ and from (30) asymptotic stability of $V$ and then $z$ is guaranteed. ∎

*Remark 4:* Replacing (23) and (24) in (21) yields to the following error dynamics

$$\dot{z} = -|f|\frac{\partial k}{\partial y}\begin{bmatrix} I_q & -BM^\top \\ \Gamma MB & \Gamma MBM^\top \end{bmatrix} z, \tag{31}$$

in which the effect of overshoot induced by $\psi^\top = -BM^\top$ can be reduced arbitrarily by tuning $\Gamma$.

*Remark 5:* If $z_2 = \theta - \zeta_2 - \beta_2$ converges to zero, an asymptotic estimate of $\theta$ is given by $\hat{\theta} = \zeta_2 + \beta_2$. Also, If $\bar{\chi} = [I_q \;\; M^\top] z$ converges to zero, an asymptotic estimate of $\chi$ is given by $\hat{\chi} = [I_q \;\; M^\top](\zeta + \beta)$. In turn, with respect to the definition of $\chi = \iota x$, an estimate of unmeasurable state is given by $\hat{x} = q^{-1}\iota^\top \hat{\chi}$.

### C. Discussion

The main question arises at this method is that, for a given regressor $\varphi(y,t)$ how to fulfill the condition $\det(M) \notin \mathcal{L}_2$. The answer lies in the $M(t)$-behavior tuning. With respect to the first order differential equation (27), the behaviour of $M(t)$ consists of *transient component* and *steady state component*. These components can be affected by arbitrarily chosen of a possible output feedback, induced by the term $\partial k/\partial y$, and the gain $B$. In addition, the transient component also depends on the initial value of $M(t)$, i.e. $M(0)$. Hence, some notes are remarked following.

*Remark 6:* In order to fulfil the condition $\det(M) \notin \mathcal{L}_2$, it is reasonable to choose $M(0)$ such that $\det(M(0)) \neq 0$. However, the effect of this choice fades away over time.

*Remark 7:* The equation (27) becomes a simpler differential equation if the positive definite matrix $B$ becomes diagonal. In this case, (27) can be decomposed to $q$ individual differential equations. Now, to satisfy the state $\det(M) \notin \mathcal{L}_2$, a necessary condition is that the entries on the main diagonal of $B$ be different. This yields different steady state components for each $q$ individual differential equation.

Generally, a necessary condition for *any* positive definite matrix $B$ in (27) is that its eigenvalues be different.

*Remark 8:* Notice that, a sudden change in the system parameters distorts the output signal in (1), (2). This can be transferred to the $M(t)$-dynamics in (27) by an output feedback which appears in the term $\partial k/\partial y$. Then, the dynamics are actuated to adapt $M(t)$ with the changes, even if the regressor $\varphi(.)$ is not dependent on the output signal, i.e. $\varphi(t)$.

## IV. SIMULATION EXAMPLE

Consider the system (1), (2) with $f = 1, g_0 = g_1 = -y, \theta = \text{col}[-1,1]$, and $\varphi^\top = [1, \varphi_2(t)]$. Assume that $\varphi_2(t): \mathbb{R}_+ \to \mathbb{R}$ is derived from the following equations

$$\varphi_2(t) = \dot{d}_1(t) + a(1+b_1)d_1(t), \tag{32}$$

$$a(b_1 - b_2)d_1(t) = \dot{d}(t) + a(1+b_2)d(t), \tag{33}$$

where $a, b_1, b_2 \in \mathbb{R}_{>0}$ are positive constant and $b_1 \neq b_2$, and $d(t)$ belongs to the following set of differentiable function

$$\begin{aligned}\mathcal{D} := \{d: \mathbb{R}_+ &\to \mathbb{R} | \dot{d}(t) \notin \mathcal{L}_2, d^{(i)}(t) \in \mathcal{L}_\infty, \\ &\lim_{t \to \infty} d^{(i)}(t) = 0, \forall i = 0,1,2\}.\end{aligned} \tag{34}$$

The set $\mathcal{D}$ is borrowed from [1] with some modifications. For any $d \in \mathcal{D}$, the condition $d^{(i)}(t) \in \mathcal{L}_\infty$ insures $\varphi_2(t)$ and then $\varphi(t)$ remains bounded. Also, the condition $\lim_{t \to \infty} d^{(i)}(t) = 0$ yields $\varphi_2(t)$ and then $\varphi(t)$ is not PE because it converges to zero as $t \to \infty$.

Let us assume $\partial k/\partial y = a$ in (12) and (27). Then, the steady state component of (12)-solution for $B = b_1 I_2$ and $B = b_2 I_2$ can be obtained as

$$\mu_{1ss}^\top = \left[\frac{1}{a(1+b_1)}, d_1(t)\right], \tag{35}$$

$$\mu_{2ss}^\top = \left[\frac{1}{a(1+b_2)}, d_1(t) + d(t)\right], \tag{36}$$

respectively. With respect to Remark 7, choosing $B = \text{diag}(b_1, b_2)$ for (27) yields to

$$M_{ss} = [\mu_{1ss}, \mu_{2ss}] \tag{37}$$

where $M_{ss}$ is the steady state component of (27)-solution.

*Remark 9:* The regressors $\mu_{1ss} \notin \text{PE}, \mu_{2ss} \notin \text{PE}$ because $\lim_{t \to \infty} d(t) = \lim_{t \to \infty} d_1(t) = 0$, and $\det(\mu_{1ss}\mu_{1ss}^\top) = \det(\mu_{2ss}\mu_{2ss}^\top) = 0$ due to positive semi definite property of the matrixes. Notice that, the matrix $M_{ss} \notin \text{PE}$, but $\det(M_{ss}) \notin \mathcal{L}_2$ because

$$\det(M_{ss}) = \frac{-d(t)}{(1+b_1)(1+b_2)} \tag{38}$$

in which, by (34), $\dot{d}(t) \notin \mathcal{L}_2$.

### A. Simulation Results

Consider the system introduced in this section with

$$d(t) = \frac{\sin t}{\sqrt{1+t}} \quad (39)$$

which $d \in \mathcal{D}$ [1]. Simulations are run to evaluate the performance of estimators introduced in Proposition 1 and 2. All initial values correspond to the system and the estimators are set to zero.

The time histories of the parameters estimation errors $\theta - \hat{\theta}$ and the state estimation errors $x - \hat{x}$ are shown for estimator $b_1$ (dashed line), estimator $b_2$ (dashed-dot line), and estimator B (solid-line) in Fig. 1. The estimator $b_i$ is formed by replacing $B = b_i I_2$ in Proposition 1 and the estimator B is formed by replacing $B = \text{diag}(b_1, b_2)$ in Proposition 2, while the mapping $k(y) = ay$ and the gain $\Gamma = I_2$ are chosen for all of them. The graphs in Fig. 1 show the performance of the estimators for the set $(a, b_1, b_2) = (0.5, 0.5, 2)$.

From the graphs in Fig. 1 it can be seen that, for $a = 0.5$, a bigger value of $b_i$ has a better response in the parameter estimation. In the all graphs, the estimator B has a faster response in parameter and state estimation. The last graph of Fig. 1 shows the absolute value of the $M(t)$-determinant. From (38) and (39) the determinant of $M(t)$ is not square integrable which is the core of convergence property of the estimator B.

The next simulation is run to show the influence of the gain $\Gamma = \gamma I_2$ on the performance of the estimator B. All conditions and gains are the same as previous simulation. The time histories of the parameters and the state estimation errors are shown for $\gamma = 1$ (dashed line), $\gamma = 100$ (dashed-dot line), and $\gamma = 10000$ (solid-line) in Fig. 2. As it is evident in Fig. 2, the overshot is decreased by increasing the value of $\Gamma$, and as a result, faster responses in the both parameter and state estimation are obtained.

## V. Conclusion

A new I&I based filtered transformation is introduced via dynamic matrix for a class of second-order systems with single-output. The main contribution of this paper is to propose an estimator based on the new transformation, which asymptotically estimates unavailable state and unknown parameters. Parameter convergence is guaranteed under non-square-integrability assumption for the matrix. Moreover, it is shown that the estimator provides an arbitrary convergence speed. Also simulation results are concluded for a class of regressors which are not PE but satisfy the new condition. Some practical results on electrical systems are under way to show effectiveness of the proposed design.

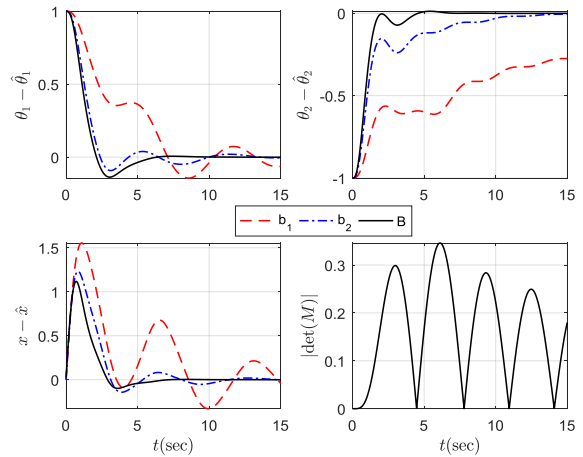

Figure 1. Transient performance of the estimator $b_i$ and estimator B.

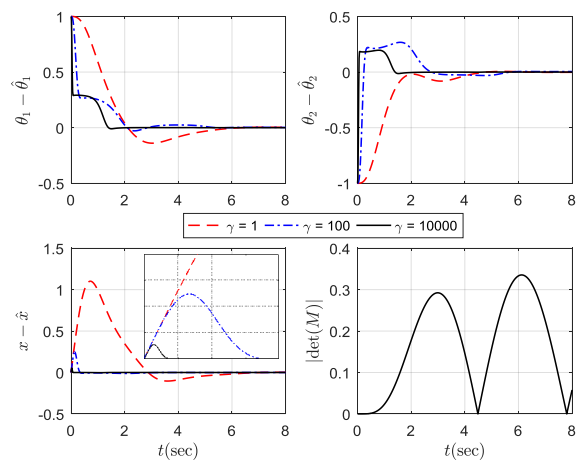

Figure 2. Transient responces of the estimator B for changs in $\gamma$.